\documentclass[lettersize,journal]{IEEEtran}

\usepackage{enumerate}
\usepackage{amsmath,amsfonts}
\usepackage[linesnumbered,ruled,vlined]{algorithm2e}

\usepackage{array}
\usepackage{textcomp}
\usepackage{url}
\usepackage{verbatim}
\usepackage{graphicx}
\usepackage{cite}
\usepackage{xcolor}
\usepackage{blindtext}
\usepackage{xparse}
\usepackage{dialogue}
\usepackage{longtable}
\usepackage{tabularx}
\usepackage{amssymb}% http://ctan.org/pkg/amssymb
\usepackage{pifont}% http://ctan.org/pkg/pifont
\newcommand{\cmark}{\ding{51}}%
\newcommand{\xmark}{\ding{55}}%

\usepackage{tikz,lipsum,lmodern}
\usepackage[most]{tcolorbox}
\usepackage{colortbl}

\DeclareDocumentCommand\dia{ o m }{%
    \begin{itemize}[%
        ,label=\IfNoValueTF {#1} {}{#1:}
        ,font=\color{black}
        ]
        \item #2
    \end{itemize}%  
    }

\usepackage{multirow}
\usepackage{comment}
\usepackage{tikz}
\usetikzlibrary{shapes.geometric, arrows}
\usepackage{subcaption}
\usepackage{graphicx}
\usepackage{float}
\usepackage{subfloat}
\usepackage{ragged2e}

\usepackage{dblfloatfix} 
\usepackage{amsmath}
\usepackage[ruled,linesnumbered]{algorithm2e}
\usepackage[noend]{algpseudocode}
\usepackage{threeparttable}

\usepackage{listings}

\definecolor{codered}{rgb}{0.8,0,0}
\definecolor{codegreen}{rgb}{0,0.6,0}
\definecolor{codegray}{rgb}{0.5,0.5,0.5}
\definecolor{codepurple}{rgb}{0.58,0,0.82}
\definecolor{backcolour}{rgb}{0.95,0.95,0.92}
\definecolor{mycolor}{rgb}{0.90,0.95,0.90}

\lstdefinestyle{ChatGPTResponses}{
    backgroundcolor=\color{mycolor},   
    commentstyle=\color{codegreen},
    keywordstyle=\color{magenta},
    numberstyle=\tiny\color{codegray},
    stringstyle=\color{codepurple},
    basicstyle=\ttfamily\footnotesize,
    breakatwhitespace=false,         
    breaklines=true,                 
    captionpos=b,                    
    keepspaces=true,   
    numbers=none,                      
    showspaces=false,                
    showstringspaces=false,
    showtabs=false,                  
    tabsize=2,
    frame=tb % draw a frame at the top and bottom of the code block
}

\lstdefinestyle{custom}{
    backgroundcolor=\color{backcolour},   
    commentstyle=\color{codegreen},
    keywordstyle=\color{codered},
    numberstyle=\tiny\color{codegray},
    stringstyle=\color{codepurple},
    basicstyle=\ttfamily\footnotesize,
    breakatwhitespace=false,         
    breaklines=true,                 
    captionpos=b,                    
    keepspaces=true,                 
    numbers=none,                    
    numbersep=5pt,                  
    showspaces=false,                
    showstringspaces=false,
    showtabs=false,                  
    tabsize=2,
    frame=tb % draw a frame at the top and bottom of the code block
}

\newcommand{\mycomment}[1]{}

\newcommand{\divas}{\texttt{DIVAS}}

\SetKwInput{KwInput}{Input}                % Set the Input
\SetKwInput{KwOutput}{Output}              % set the Output

\begin{document}

\title{DIVAS: An LLM-based End-to-End Framework for SoC Security Analysis and Policy-based Protection}

\author{Sudipta Paria,~\IEEEmembership{Student Member, IEEE},
Aritra Dasgupta,~\IEEEmembership{Student Member, IEEE} \\
Swarup Bhunia,~\IEEEmembership{Senior Member, IEEE} \\
Department of Electrical and Computer Engineering, University of Florida, Gainesville, Florida
        % <-this % stops a space
%\thanks{}% <-this % stops a space
%\thanks{}
}

% The paper headers
\markboth{}%
{Shell \MakeLowercase{\textit{et al.}}: }

\maketitle

\begin{abstract}

Securing critical assets in a bus-based System-On-Chip (SoC) is imperative to mitigate potential vulnerabilities and prevent unauthorized access, ensuring the system's integrity, availability, and confidentiality. Ensuring security throughout the SoC design process is a formidable task owing to the inherent intricacies in SoC designs and the dispersion of assets across diverse IPs. Large Language Models (LLMs), exemplified by OpenAI's ChatGPT and Google BARD, have showcased remarkable proficiency across various domains, including security vulnerability detection and prevention in SoC designs. In this work, we propose \divas, a novel framework that leverages the knowledge base of LLMs to identify security vulnerabilities from user-defined SoC specifications, map them to the relevant Common Weakness Enumerations (CWEs), followed by the generation of equivalent assertions, and employ security measures through enforcement of security policies. The proposed framework is implemented using multiple ChatGPT and BARD models, and their performance was analyzed while generating relevant CWEs from the SoC specifications provided. The experimental results obtained from open-source SoC benchmarks demonstrate the efficacy of our proposed framework.

\end{abstract}

\begin{IEEEkeywords}
SoC Security, Security Policies, Assertion Based Verification, CWEs, ChatGPT, BARD
\end{IEEEkeywords}

\section{Introduction}

\begin{figure}[t]
\centering
\includegraphics[scale=0.45]{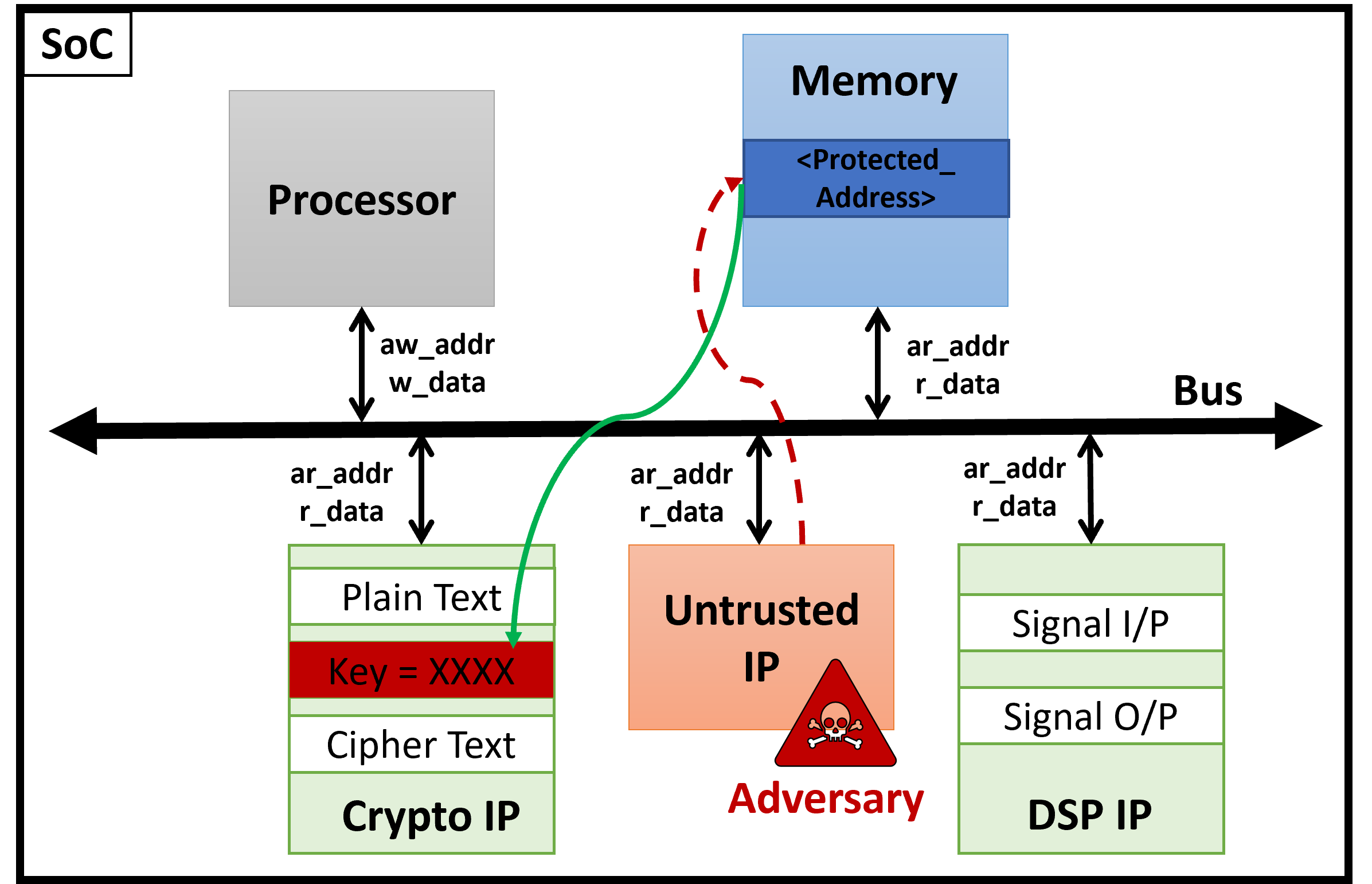}
\caption{A model SoC framework with Trusted and Untrusted IPs and their interactions.}
\label{fig:example_soc}
\end{figure}

A bus-based System-On-Chip (SoC) integrates multiple functional components into a single chip and utilizes a common bus to facilitate communication between these components. The globalization of the IC supply chain has forced the semiconductor industry to adopt a Zero Trust model for security. Under this model, malicious entities can exploit the vulnerabilities at any stage of the design flow. Hence, it is essential to incorporate preventive measures to protect the secure assets in an SoC from potential threats. Fig. \ref{fig:example_soc} depicts an illustrative example using a model SoC to emphasize the importance and criticality of secure information and authorized access to assets of the SoC. Security policies offer actionable specifications for SoC designers, architects, and security experts by instantiating Confidentiality, Integrity, and Availability (CIA) requirements for certain assets. The growing intricacy of SoC designs has made it challenging for developers to identify and fix vulnerabilities in complex SoC designs. Despite considerable efforts to ensure functional accuracy in hardware designs, there needs to be more focus on automation frameworks to identify security vulnerabilities and enforce security requirements. Fixing vulnerabilities in the later stages of the design flow can be extremely difficult and expensive.

In recent years, the collaborative endeavors of the semiconductor industry have introduced hardware-related concerns to the list of Common Weakness Enumerations (CWEs) maintained by the MITRE Corporation \cite{CWE}. CWEs serve as a ``common language'' for identifying and mapping vulnerabilities. Identifying distinct vulnerabilities necessitates varying degrees of understanding regarding design, secure assets, the threat model, security requirements, etc. \cite{hardfails, hunt_bugs}. The existing approaches primarily involve manual assessment of the hardware description language (HDL) code which heavily relies on human expertise and experience that might not be sufficient to discover all potential vulnerabilities. Latest works on bug fixing in hardware designs involve generic repair templates \cite{repair_book, repair_paper_1}, using static analysis and security-related feedback \cite{cweat_paper}, Genetic Programming \cite{cirfix_paper}, and Large Language Models (LLMs) \cite{llm_bug_fix_paper, llmpaper_2}. However, the current methodologies are limited to hardware designs, mainly covering specific vulnerabilities but not explicitly addressing security requirements for generic bus-based SoC designs. 

\begin{figure*}[t]
\centering
\includegraphics[scale=0.45]{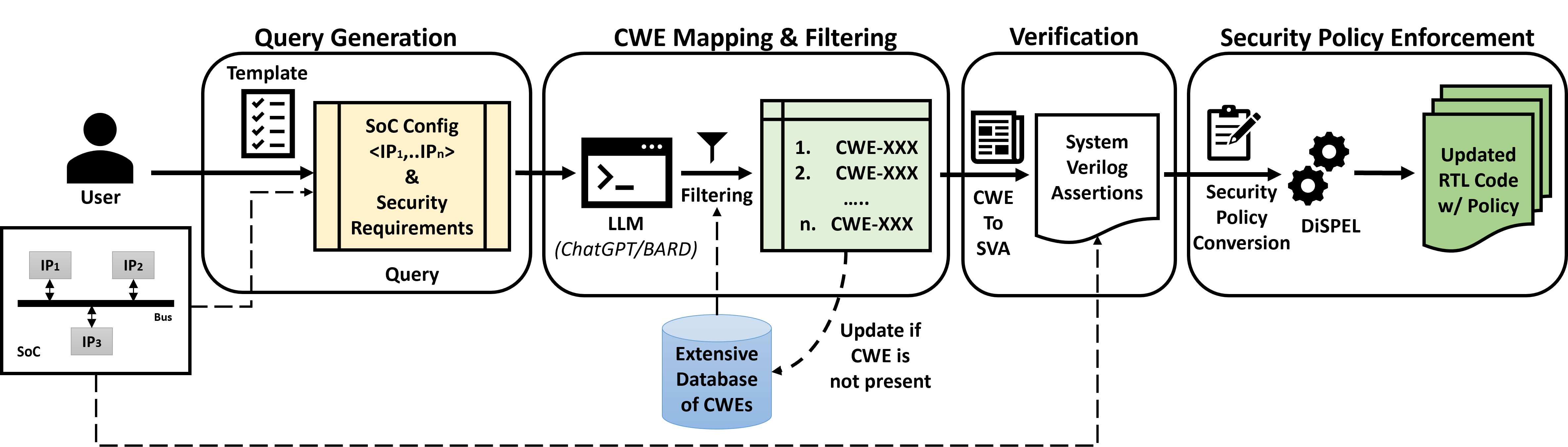}
\caption{\divas: Overview of the proposed framework.}
\label{block_diagram}
\end{figure*}

In this paper, we propose \divas, an LLM-based end-to-end framework for SoC security analysis and policy-based protection. \divas~leverages the ability of LLMs to identify the CWEs for a given SoC specification and employs a novel LLM-based filtering technique to determine the relevant CWEs, which are then converted into SystemVerilog Assertions (SVAs) using LLMs for verification. The proposed methodology generates 3-tuple security policies from these SVAs using DiSPEL \cite{dispel_arxiv}, an automated tool flow to parse the security policies and generate RTL code for enforcing policies. The DiSPEL tool enforces these policies through a centralized security module across the bus interconnect for bus-level security policies or by appending the RTL code block in the top-level wrapper that interacts with the bus interface for IP-level security policies. \divas~provides an automated framework that can take user specifications for any generic SoC, identify relevant CWEs using LLMs, generate SVAs for verification, create corresponding security policies, and incorporate the translated policies through a security module or wrapper. The proposed \divas~framework is illustrated in Fig. \ref{block_diagram}.

In summary, this paper presents the following contributions:
\begin{itemize}
    \item An automation framework for generating queries from user-given specifications and security requirements to identify CWEs by leveraging the LLM knowledge base. 
    \item Curating an extensive list of CWEs with different classifications for bus-level and IP-level vulnerabilities and incorporating a filtering methodology to retain only relevant CWEs for a given SoC context. 
    \item Analyzing and correcting SVAs generated by LLMs for employing simulation-based validation and formal verification.
    \item Converting SVAs to respective security policies in a 3-tuple format representation and employing DiSPEL tool for enforcing security policies through a centralized security module or bus-level wrapper as required.
\end{itemize}

The remainder of this paper is organized as follows: Section II provides the relevant background and describes the motivation behind our work. Section III outlines the overall flow of our proposed framework with a brief description of each stage. Section IV contains the experimental results and analysis. Lastly, we conclude and provide future directions in Section V.

\section{Background}

\subsection{Security Requirements in Bus-based SoC}
A bus-based SoC refers to a type of SoC architecture where the different components of the SoC are connected via a bus interconnect. The bus interconnect provides a communication channel between different IP cores, memory blocks, and peripheral devices of the SoC. Bus-based SoCs are complex systems that contain multiple IP cores from different vendors. These IP cores may have different levels of security and trustworthiness, making them vulnerable to attacks. Integrating these IP cores in the SoC can create new security vulnerabilities, which need to be identified and mitigated to ensure the system's overall security. SoCs are used in various applications, including mobile devices, wearables, smart homes, medical devices, and automotive systems. These applications often handle sensitive and confidential data, such as personal information, financial data, and medical records. Therefore, securing SoC assets is crucial to protect sensitive data, prevent unauthorized access, theft, or manipulation, and ensure the reliable operation of the system. It involves implementing security mechanisms and protocols that can detect and prevent security threats, limit the access of untrusted entities, and provide secure communication channels between different system components.

\subsection{Common Weakness Enumerations}

The list of CWEs from MITRE Corporation \cite{CWE} provides a collection of common software and hardware weaknesses that may cause potential security vulnerabilities maintained by the CWE community. A software, firmware, or hardware design bug is considered a “weakness” if it can be exploited under any circumstances. The CWEs and associated classification taxonomy serve as a common language that can be used to identify and describe these weaknesses in terms of CWEs. These weaknesses can be recognized and discussed in terms of CWEs using the CWE-ID and its related classification taxonomy. The list of CWEs is widely adopted in the industry for verification purposes to ensure the design is resistant to common vulnerabilities. The current industry practices involve experienced security experts responsible for testing the design against the most common CWEs for the respective IPs. The most commonly adopted verification methodologies are: (i) assertion-based validations and (ii) writing security properties or policies. Both of these methodologies rely on the expertise and manual efforts of the security expert personnel.

Assertion-Based Validation (ABV) is widely used in the industry for functional verification in bus-based SoCs. Assertions are condition-based validation checks to identify any particular event occurrence and generate appropriate warnings or messages if the condition is met. For example, we can think of an assertion that verifies the output of an adder is always equal to the sum of its input or not. In addition to  functional validation, assertions are required to detect security vulnerabilities in SoCs. Such security assertions are intended to identify any deviation from the security specifications for a given SoC and identify potential vulnerabilities which may lead to future attacks if not fixed. Once the vulnerability is identified for an IP in the SoC under test, the immediate next step would be fixing the respective IP module. Simulation-based validation and formal verification methods can help identify the vulnerabilities in the system design by activating assertions. However, such methodologies are insufficient due to limited scope, lack of scalability, and design complexity, which originates the need to incorporate security policies through which the security requirements are represented formally.     

A security policy for a SoC is a set of guidelines, rules, and procedures that define the security requirements and protection mechanisms for the SoC. The security policy aims to ensure the SoC's confidentiality, integrity, availability, and data and functionality. The security policy should address various aspects of SoC design, including hardware, software, and communication interfaces. As part of the design process, most SoC design specifications include security policies that define access constraints to the sensitive data or assets that must be protected. The policies are defined across multiple design planning and development stages and then updated or distilled across development and validation stages. However, the process of identifying relevant vulnerabilities, defining security requirements through policies, and subsequent enforcement remains exceedingly intricate and predominantly reliant on manual efforts in current practices.\\

\noindent\textbf{Example of Security Policy}

The SoC model depicted in Fig. \ref{fig:example_soc} consists of a Master IP (the processor core) and several $3^{rd}$ party Slave IPs like the crypto, DSP, and memory, as well as peripheral IPs for external communication through SPI. The SoC is designed to function in a way that restricts access to the secure memory address of on-chip RAM to only trusted IPs. The crypto IP reads the plaintext, encrypts it using stored keys, and then stores the ciphertext in a trusted memory region that other trusted IPs can access. 

A security policy can be implemented to achieve access control by limiting the access of the bus by untrusted IPs to prevent the unauthorized disclosure of private keys while in transit from memory to crypto IP.  Considering the SoC uses AXI4 bus protocol, the bus signals are represented as \textit{aw\_addr}, which contains the address where the data is to be written, and \textit{w\_data}, which contains the data to be written. The security requirement can be achieved by restricting the bus access for the untrusted IP while sensitive data is transmitted through the shared bus. Listing \ref{lst:1} depicts how to implement this security policy using a centralized security module. 

\lstset{style=custom}
\begin{lstlisting}[caption=Security Policy Example, language=Verilog, label=lst:1]
if(slave[`Crypto'].aw_addr >= 32'h93000014 
&& slave[`Crypto'].aw_addr <= 32'h93000028)
{
    slave[`SPI'].w_data = 32'h0;
}
\end{lstlisting}

\begin{table*}[t]
\centering
\begin{threeparttable}
\caption{Comparison of \divas with existing solutions}
\label{tab:comparison}
 \begin{tabular}{ccccccc} 
 \hline %\hline
\multirow{2}*{\textbf{Proposed Solutions}}  & \textbf{Applicable}  & \textbf{Uses} & \textbf{Mapping}  & \textbf{Support for}  & \textbf{Generates}  & \textbf{Performs}  \\ 
  & \textbf{to SoCs?} & \textbf{LLMs?} & \textbf{to CWEs?} & \textbf{Generic SoCs?} & \textbf{Assertions}? & \textbf{Code Fix?} \\ 
\hline
CirFix \cite{cirfix_paper} & \textcolor{red}{\xmark} & \textcolor{red}{\xmark} & \textcolor{red}{\xmark} & \textcolor{red}{\xmark} & \textcolor{red}{\xmark} & \cmark  \\ %cirfix
Don’t CWEAT It \cite{cweat_paper} & \textcolor{red}{\xmark} & \textcolor{red}{\xmark} & \cmark & \textcolor{red}{\xmark} & \textcolor{red}{\xmark} & \cmark   \\ %Don't CWEAT It
Chip-Chat \cite{chipchat_paper} & \textcolor{red}{\xmark} & \cmark & \textcolor{red}{\xmark} & \textcolor{red}{\xmark} & \textcolor{red}{\xmark} & \textcolor{red}{\xmark}   \\ % ChipChat
B. Ahmad \emph{et al.} \cite{llm_bug_fix_paper} & \cmark  & \cmark  & \cmark & \textcolor{red}{\xmark} & \textcolor{red}{\xmark} & \cmark   \\ %LLM based HW Bug fix 
R. Kande \emph{et al.} \cite{llmpaper_2} & \cmark  & \cmark  & \cmark  & \textcolor{red}{\xmark} & \cmark  & \textcolor{red}{\xmark}  \\ %LLM based Assertion
\divas* & \cmark  & \cmark  & \cmark  & \cmark  & \cmark & \cmark  \\
 \hline
\end{tabular} 
%\noindent \\\footnotesize{ * }
\begin{tablenotes}
    \item[*] {\footnotesize current work}
\end{tablenotes}
\end{threeparttable}
\end{table*}

\subsubsection{Common CWE(s) in Hardware}
Various commercial and community tools offer static analysis for HDL to detect errors and bugs. However, they offer limited capability since these tools primarily focus on functional and structural checks only \cite{formal_book_cheri}, and they do not address the issue of design security that may be functionally correct but insecure. The search space for exploring potential vulnerabilities for any SoC design with multiple IPs is huge, considering the complex design, limited access, interaction between components, and other constraints. This initiates a requirement for a standardized way of identifying and describing relevant security concerns from the design specifications for an SoC design. The introduction of CWEs has helped to standardize and simplify the process of identifying and mitigating security vulnerabilities in SoCs. CWEs are widely adopted by security researchers and system developers to identify and classify different types of security vulnerabilities or weaknesses in H/W designs and deploy necessary countermeasures.

The following CWEs are examples of some well-known security vulnerabilities for any bus-based SoC:
\begin{itemize}
    \item \textbf{CWE-284:} Improper Access Control refers to a weakness where a system does not properly restrict access to its resources or operations. In the context of a bus-based SoC security, this can manifest as inadequate protection of access to sensitive resources, such as memory regions or peripherals, when using a shared bus for communication between components within the SoC.
    \item \textbf{CWE-522:} Insufficiently Protected Credentials refer to a weakness where a system does not adequately protect sensitive data, such as passwords, cryptographic keys, or other authentication information. In the context of a bus-based SoC security, this can manifest as inadequate protection or handling of sensitive data within the SoC during storage, processing, or transmission over the bus.
    \item \textbf{CWE-1245:} Improper Finite State Machines (FSMs) in Hardware Logic refers to a weakness where an incomplete or incorrect implementation of FSMs allows an attacker to put the system in an undefined state. In the context of a bus-based SoC security, this can manifest as an inadequate error-detection mechanism that allows the attacker to cause a Denial of Service (DoS) or gain privileges on the victim's system.
    \item \textbf{CWE-1231:} Improper Prevention of Lock Bit Modification refers to a weakness where the system does not prevent the value of the lock bit from being modified after it has been set. In the context of a bus-based SoC security, this can manifest as inadequate protection on a trusted lock bit for restricting access to registers, address regions, or other resources. 
\end{itemize}

\subsection{Large Language Models}

The development of AI has led to many breakthroughs in Natural Language Processing (NLP), which the industry has widely adopted due to the increasing demand for conversational models. Over the years, LLMs have consistently showcased impressive performance across various NLP tasks. Pre-trained transformer models have shown remarkable efficacy in identifying relevant bugs or vulnerabilities from informal or unstructured natural language descriptions. The evolution in LLMs is evident in the continuous evolution of highly capable models like BERT (Bidirectional Encoder Representations from Transformers) \cite{bert}, GPT-2 (Generative Pre-trained Transformer 2) \cite{gpt-2}, GPT-3 (Generative Pre-trained Transformer 3)\cite{gpt-3}, RoBERTa (A Robustly Optimized BERT Pretraining Approach) \cite{roberta}, etc. to name a few. In this work, we have explored the two most popular competing LLMs: ChatGPT by OpenAI \cite{chatgpt} and BARD by Google \cite{bard}. 

\subsubsection{Conversational LLMs}
ChatGPT is a well-known AI chatbot developed by OpenAI, which is built on top of LLMs and fine-tuned using both supervised and reinforcement learning techniques. ChatGPT was launched as a prototype in Nov 2022 and generated huge interest and attention worldwide due to its performance across many knowledge domains. It is a transformer-based neural network pre-trained with over 175 billion parameters and huge quantities of text data until Sep 2021. It makes it capable of inferring relationships between words within the text and generating context-sensitive responses. On the contrary, BARD is built on Pathways Language Model 2 (PaLM 2), a language model released in late 2022 preceded by LaMDA, which is short for Language Model for Dialogue Applications. These language models were built by fine-tuning a family of Transformer-based neural language models, an open-source neural network architecture originally developed by Google. BARD is believed to have been trained on 137 billion parameters and 1.56 million words of public dialog data and web text, though Google didn't officially disclose the exact numbers. Unlike ChatGPT, BARD can pull from the data available on the internet today. While both are built on Transformers with billions of parameters to fine-tune the model and overlapping training data sources, their performance varies with use cases, and both have their own set of limitations.

\subsubsection{Using LLMs for Security Automation}
The responses and the capability of articulating answers to different queries by ChatGPT and BARD were very promising and impressively detailed, even though they lack factual accuracy in many domains and have a limited knowledge base. We have also explored the capability of ChatGPT and BARD models to produce factual answering to domain-specific queries related to SoC security. The main goal is to identify or map the user-given design specifications and security requirements to the existing list of CWEs available on the web using the knowledge base of LLMs like ChatGPT and BARD. We have comprehensively analyzed the performance of ChatGPT and BARD models under various attack models prevalent in the SoC security literature. This includes but is not limited to bus-based attacks, side-channel attacks, timing attacks, access control violations, etc.

\subsection{Related Works}
%\subsubsection{SoC Security Evaluation}
Various techniques can bolster the security of hardware designs, like adopting a Security Development Lifecycle (SDL) that operates concurrently with the standard development procedure. The SDL involves several stages, starting from {\em `planning'}, where security requirements are determined, to {\em`architecture'} and {\em`design'}, where relevant threat models are considered. The design is reviewed using security threat models in the {\em`implementation'} and {\em`verification'} stages, with manual checks and static code analysis in the {\em`implementation'} phase. The bulk of validation is carried out through security properties expressed as assertions in HDLs during the {\em`verification'} stage. Physical testing is conducted after the fabrication to identify any vulnerabilities that may still be present. 

Several techniques have been proposed for security analysis across the SDL, such as Formal Verification \cite{formal_ver_1, formal_ver_2, formal_ver_3, formal_ver_4}
, Information Flow Tracking \cite{info_flow_paper_1, info_flow_paper_2, info_flow_paper_3, info_flow_paper_4}
, Fuzz Testing \cite{fuzzing_paper_1, fuzzing_paper_2, fuzzing_paper_3}
, and Run-time Detection \cite{runtime_paper_1, runtime_paper_2, runtime_paper_3}. Previous work on security analysis using assertion-based verification \cite{soc_sec_book, assertion_paper_1, assertion_paper_2, assertion_paper_3} focused on specific SoC designs. These techniques are either simulation-based or operate in the field with complete or near-complete designs. The identification and deployment of necessary fixes must be made earlier (during the {\em `implementation'} and {\em `verification'} phases) to prevent them from propagating to the following stages. 

Generic code repair templates are available for fixing bugs in hardware designs \cite{repair_book, repair_paper_1}. CirFix \cite{cirfix_paper} was proposed as a framework for automatically repairing defects in hardware designs using Genetic Programming. Ahmad et al. \cite{cweat_paper} explored the CWEs related to H/W designs and provided security-related feedback using static analysis to identify security bugs at the early stage of development. In the realm of software bug fixes, the software domain delves into the utilization of machine learning-driven methods like Neural Machine Translation \cite{ml_bug_fix_1} and pre-trained transformers \cite{ml_bug_fix_1}. Pearce et al. \cite{repair_1} employed a similar strategy to rectify instances of security vulnerabilities in Verilog code, successfully addressing two distinct scenarios. Ahmad et al. \cite{llm_bug_fix_paper} devised a framework for repairing specific hardware security bugs using OpenAI Codex and CodeGen LLMs. Kande et al. \cite{llmpaper_2} developed a prompt-based assertion generation framework using LLMs for a set of SoC benchmarks but limited to only the generation and correctness evaluation of assertions for a fixed set of SoC benchmarks. Existing techniques are mostly restricted to specific SoC benchmarks and do not fully automate the identification of the relevant vulnerabilities in terms of CWEs and fixing them. Table \ref{tab:comparison} provides a comprehensive overview of the existing solutions and demonstrates the necessity of our proposed framework.

\begin{table*}[t]
\centering
\caption{List of CWEs generated by ChatGPT and BARD under various threat models}
\label{tab:cwe_attack_models}
%\resizebox{\columnwidth}{!}{%
\begin{tabular}{|p{0.2\linewidth} | p{0.37\linewidth}| p{0.37\linewidth}|}
\hline
\multicolumn{1}{|c|}{\textbf{Attack Assumptions}}       & \multicolumn{1}{|c|}{\textbf{CWEs generated by ChatGPT}}        & \multicolumn{1}{|c|}{\textbf{CWEs generated  by BARD}}                                                        \\ \hline
\multicolumn{1}{|c|}{Bus-Based Attacks} & \begin{tabular}{p{0.95\linewidth}}CWE-120: Buffer Copy without Checking Size of Input('Classic Buffer Overflow')\\ CWE-121: Stack-based Buffer Overflow\\ CWE-125: Out-of-bounds Read\\ CWE-134: Use of Externally-Controlled Format String\\ CWE-352: Cross-Site Request Forgery (CSRF)\end{tabular} & \begin{tabular}{p{0.95\linewidth}}CWE-506: Timing-based information disclosure \\CWE-713: Information Exposure Through Timing Channels: Microarchitectural Data Sampling (MDS) \\CWE-905: Information Exposure Through Timing Channels: Branch Prediction \\CWE-135: Sensitive Data Exposure Through Timing Channels \\CWE-196: Power Analysis Attack (Data or Control Flow)\end{tabular} \\ \hline
\multicolumn{1}{|c|}{Side-Channel Attacks} & \begin{tabular}{p{0.95\linewidth}}CWE-319: Cleartext Transmission of Sensitive Information\\ CWE-311: Missing Encryption of Sensitive Data\\ CWE-200: Information Exposure\\ CWE-327: Use of a Broken or Risky Cryptographic Algorithm\\ CWE-769: Inefficient Algorithm\end{tabular} & \begin{tabular}{p{0.95\linewidth}}CWE-126: Time-of-check/time-of-use (TOCTOU) race condition \\CWE-135: Sensitive Data Exposure Through Timing Channels \\CWE-196: Power Analysis Attack (Data or Control Flow) \\CWE-327: Cache Timing Attack \\CWE-320: Information exposure through timing\end{tabular} \\ \hline
\multicolumn{1}{|c|}{Timing Attacks} & \begin{tabular}{p{0.95\linewidth}}CWE-200: Information Exposure\\ CWE-208: Observable Timing Discrepancy\\ CWE-710: Improper Adherence to Coding Standards\\ CWE-751: Improper Use of Platform Timer\\ CWE-613: Insufficient Session Expiration\end{tabular} & \begin{tabular}{p{0.95\linewidth}}CWE-135: Sensitive Data Exposure Through Timing Channels \\CWE-327: Cache Timing Attack \\CWE-506: Timing-based information disclosure \\CWE-713: Information Exposure Through Timing Channels: Microarchitectural Data Sampling (MDS) \\CWE-905: Information Exposure Through Timing Channels: Branch Prediction\end{tabular} \\ \hline
\multicolumn{1}{|c|}{DoS Attacks} & \begin{tabular}{p{0.95\linewidth}}CWE-400: Uncontrolled Resource Consumption ('Resource Exhaustion') - IP Level \\CWE-494: Download of Code Without Integrity Check \\CWE-613: Insufficient Session Expiration \\CWE-400: Uncontrolled Resource Consumption ('Resource Exhaustion') - Bus Level \\CWE-693: Protection Mechanism Failure\end{tabular} & \begin{tabular}{p{0.95\linewidth}}CWE-321: Resource Consumption (CPU Time, Memory, Disk Space) \\CWE-509: Insufficient Boundary Checking \\CWE-601: Buffer Overrun \\CWE-704: Improper Synchronization \\CWE-754: Improper Validation of Cryptographic Inputs\end{tabular} \\ \hline
\multicolumn{1}{|c|}{Confidentiality Attacks} & \begin{tabular}{p{0.95\linewidth}}CWE-200: Information Exposure \\CWE-311: Missing Encryption of Sensitive Data \\CWE-327: Use of a Broken or Risky Cryptographic Algorithm \\CWE-200: Information Exposure \\CWE-310: Cryptographic Issues \\\end{tabular} & \begin{tabular}{p{0.95\linewidth}}CWE-119: Improper input validation \\CWE-125: Buffer overflow \\CWE-200: Information Exposure Through Improper Error Handling \\CWE-201: Information Exposure Through Shared Resources \\CWE-284: Improper Access Control\end{tabular} \\ \hline
\multicolumn{1}{|c|}{Access Control} & \begin{tabular}{p{0.95\linewidth}}CWE-285: Improper Authorization \\CWE-306: Missing Authentication for Critical Function \\CWE-732: Insecure Permission Assignment for Critical Resource \\CWE-250: Execution with Unnecessary Privileges \\CWE-724: OWASP Top Ten 2017 Category A6: Security Misconfiguration\end{tabular} & \begin{tabular}{p{0.95\linewidth}}CWE-284: Improper Access Control \\CWE-805: Security Features Bypass \\CWE-895: Insufficiently Protected Credentials \\CWE-918: Trust Boundary Violation \\CWE-928: Insufficiently Protected Cryptography\end{tabular} \\ \hline
\end{tabular}
%}
\end{table*}

\section{Methodology/Proposed Framework}
    
    \begin{figure}[!ht]
    \centering
    \includegraphics[scale=0.6]{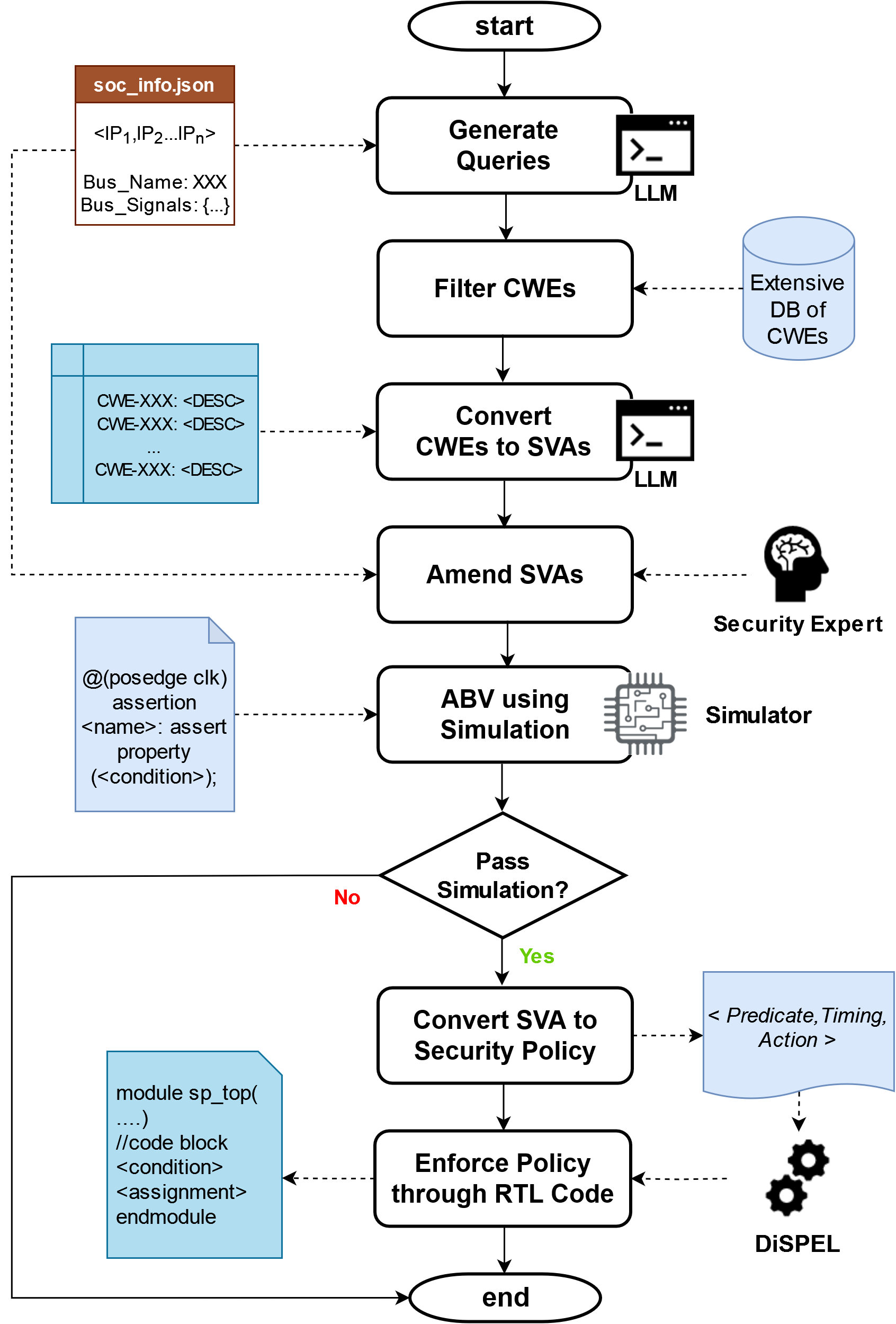}
    \caption{\divas: Flow diagram.}
    \label{flow_diagram}
    \end{figure}       
    
In this section, we describe the main components of the \divas~framework, starting from the design specifications of the SoC under test, followed by the generation \& verification of security assertions and enforcing security requirements through the implementation of synthesizable Verilog code. The overall flow has been depicted in Fig. \ref{flow_diagram}. The proposed framework can be categorized into four major stages: (I) Design Specification \& Query Generation, (II) CWE Mapping using LLMs \& Filtering, (III) Security Assertion Creation \& Verification, and (IV) Translation to Security Policy \& Policy Enforcement using RTL. We will now discuss each of these stages in detail.
 
    \subsection{Design Specification \& Query Generation}
        This marks the initial stage necessitating user engagement, succeeded by the automated generation of queries pertinent to the SoC under test. The SoC specifications are formally presented in a manner that enables \divas~to tokenize distinct details concerning IPs, bus-level configuration, and the overall SoC, facilitating the generation of corresponding queries. We have prepared a survey template in the form of a list of common questions related to available open-source SoCs for better usability and easement for a user who may or may not be knowledgeable about the exact security requirements or common vulnerabilities. The user responses are recorded and translated into a pre-defined format represented in a JSON file which serves as the design specifications for the SoC under test. The automated flow will parse the JSON file and generate query sentences using the available information about the SoC under consideration. The query will also include assumptions provided by the user, if any. The current implementation can generate relevant queries for the IPs available in any bus-based SoC design. The queries are then fed into LLMs such as ChatGPT or BARD for further steps.

\begin{table*}[t]
\centering
\caption{A snapshot of the Extensive DB of CWEs used in \divas}
\label{tab:extensive_db}
\begin{tabular}{|c| p{0.3\linewidth}|c|c|c|}
%{|c|l|l|l|l|}
\hline
\textbf{CWE\#} & \multicolumn{1}{c|}{\textbf{Bug Description}} & \textbf{Classification} & \textbf{Timing Requirements} & \textbf{Type of Violation} \\ \hline
CWE-119 & Improper   Restriction of Operations within the Bounds of a Memory Buffer & Bus + IP Level & Asynchronous & Access Control \\ \hline
CWE-120 & Buffer   Copy without Checking Size of Input ('Classic Buffer Overflow') & N/A & N/A & Inadequate Error   Handling \\ \hline
CWE-121 & Stack-based   Buffer Overflow & N/A & N/A & Inadequate Error   Handling \\ \hline
CWE-125 & Out-of-bounds   Read & Bus + IP Level & Asynchronous & Inadequate Error   Handling \\ \hline
CWE-1059 & Insufficient   Technical Documentation & N/A & N/A & Inadequate Error   Handling \\ \hline
CWE-131 & Incorrect   Calculation of Buffer Size & N/A & N/A & Inadequate Error   Handling \\ \hline
CWE-1390 & Weak   Authentication & Bus + IP Level & Synchronous & Access Control \\ \hline
CWE-1391 & Use   of Weak Credentials & IP Level & Synchronous & Access Control \\ \hline
CWE-190 & Integer   Overflow or Wraparound & N/A & N/A & Inadequate Error   Handling \\ \hline
CWE-20 & Improper   Input Validation & IP Level & Synchronous & Inadequate Error   Handling \\ \hline
CWE-200 & Exposure   of Sensitive Information to an Unauthorized Actor & Bus + IP Level & Synchronous & Information Flow \\ \hline
CWE-226 & Sensitive   Information in Resource Not Removed Before Reuse & Bus + IP Level & Synchronous & TOCTOU \\ \hline
CWE-284 & Improper   Access Control & Bus + IP Level & Synchronous & Access Control \\ \hline
CWE-285 & Improper   Authorization & Bus + IP Level & Synchronous & Access Control \\ \hline
CWE-287 & Improper   Authentication & Bus + IP Level & Synchronous & Access Control \\ \hline
CWE-310 & Cryptographic   Issues & IP Level & Synchronous & Information Flow \\ \hline
CWE-325 & Missing Required Cryptographic Step & IP Level & Asynchronous & Information Flow \\ \hline
CWE-326 & Inadequate   Encryption Strength & IP Level & Asynchronous & Information Flow \\ \hline
CWE-327 & Use   of a Broken or Risky Cryptographic Algorithm & IP Level & Asynchronous & Information Flow \\ \hline
CWE-330 & Use   of Insufficiently Random Values & IP Level & Asynchronous & Information Flow \\ \hline
CWE-362 & Concurrent   Execution using Shared Resource with Improper Synchronization & Bus Level & Synchronous & Liveness \\ \hline
CWE-367 & Time-of-check   Time-of-use (TOCTOU) Race Condition & Bus + IP Level & Synchronous & TOCTOU \\ \hline
CWE-522 & Insufficiently   Protected Credentials & Bus + IP Level & Asynchronous & Access Control \\ \hline
CWE-665 & Improper   Initialization & Bus + IP Level & Asynchronous & Information Flow \\ \hline
CWE-667 & Improper   Locking & Bus + IP Level & Synchronous & Information Flow \\ \hline
CWE-787 & Out-of-bounds   Write & Bus + IP Level & Asynchronous & Inadequate Error   Handling \\ \hline
CWE-798 & Use   of Hard-coded Credentials & Bus + IP Level & Synchronous & Information Flow \\ \hline
CWE-862 & Missing   Authorization & Bus + IP Level & Asynchronous & Access Control \\ \hline
CWE-863 & Incorrect   Authorization & Bus + IP Level & Asynchronous & Access Control \\ \hline
\end{tabular}
\end{table*}

    \subsection{CWE Mapping using LLMs \& Filtering}
        The proposed framework leverages context-sensitive factual answering capabilities of LLM, such as ChatGPT or BARD, to identify the most relevant CWE for the SoC under test. The performance of ChatGPT and BARD in identifying the relevant CWEs solely depends on the context created by some specific words or tokens in queries generated from the design specifications and user responses. We observed that the list of CWEs generated by ChatGPT and BARD varies with the user's design specifications and assumptions. Table \ref{tab:cwe_attack_models} tabulates CWEs generated by ChatGPT and BARD for different attack models. Evidently, the responses from LLMs vary with input queries at different time instances, which might be irrelevant and unreliable at times. In addition to that, both ChatGPT and BARD  also enlist some CWEs related to software-level vulnerabilities, which are not directly related to bus-based SoC design due to the limited knowledge base and training in this particular domain. Hence, it is required to employ a filtering technique to identify only the relevant CWEs from the list of CWEs generated by LLMs so that the appropriate security assertions and necessary logic for ensuring security can be deployed. We have prepared an extensive list of CWEs, referred as Extensive DB ($\Lambda$), that will be used in the filtering process. This database has been prepared after rigorously analyzing different SoC configurations and the common list of H/W and SoC level vulnerabilities available at \cite{CWE}. The classifications have been done in terms of timing requirements, positioning conditions, and the type of weakness. Table \ref{tab:extensive_db} represents a partial snapshot of the Extensive DB with different CWEs and respective classifications.  

        Let us first describe the notations used in the algorithmic steps for better understanding.
        \begin{itemize}
            \item  SoC Configuration is represented as: \\
                  $S = \{IP_1,IP_2,...IP_q\}$ \\
                  $IP_i$ = {\em\{NAME, DESC, OP, BASE, RANGE, PROC\_ADD\_R\}}

            \item Extensive DB, denoted by $\Lambda$, where,\\ $\Lambda = \{\Lambda_1,\Lambda_2,...\Lambda_p\}$ and \\
            $\Lambda_i$ = $<$ {\em CWE\_ID, DESC, BUS, IP, SYNC, TYPE, MISC} $>$ 
                  
            \item List of CWEs from LLM Responses, denoted by $\Omega$\\ where, $\Omega = \{\Omega_1,\Omega_2,...\Omega_n\}$
            
            \item Filtered List of CWEs, denoted by $\omega$\\ where, $\omega = \{\omega_1,\omega_2,...\omega_k\}$

        \end{itemize}
         
         Algorithm \ref{algo:filter_cwe} describes the steps of the filtering method for identifying only the relevant CWEs $\omega$ from the LLM-generated response $\Omega$. Filtering evaluates the relevance of each CWE for the given SoC context based on the data available in the Extensive DB ($\Lambda$). The Extensive DB, comprising about 180 CWEs and respective classifications, has been prepared after broadly studying different CWEs classified under Common Hardware Design Bugs (CWE VIEW: H/W Design (CWE-1194)). The CWEs are matched using the ID value or the description text. We used semantic context matching utilizing Cosine Similarity scores for the description text to map the CWE description text to the most relevant entry in Extensive DB. Table \ref{tab:cosine_score} shows two example description texts and some matching texts in the Extensive DB with the respective Cosine Similarity score. The text with the highest similarity score will be selected, and the automated flow will perform respective mapping for the particular CWE-ID. However, the current list of CWEs may not be complete. If any CWE-ID relevant to H/W or SoC security generated by LLMs is not present in $\Lambda$, then it can be included using the automated tool.
    
         Algorithm \ref{algo:map_cwe} describes the process of extracting the classification details for a particular CWE from the Extensive DB and mapping to the respective type based on the values. All the relevant CWEs, whether bus-level, IP-level, or both, are appended to the filtered list $\omega$.

\begin{algorithm}[!ht]
\caption{Filter CWEs}
\label{algo:filter_cwe}

\DontPrintSemicolon
\KwInput{List of CWEs from LLM Responses, denoted by $\Omega$}
\KwOutput{Filtered List of CWEs, denoted by $\omega$ }
\KwData{Extensive DB, denoted by $\Lambda$}

       \While{$\Omega$ is not Empty}
       {
       		Parse $\Omega_i \in \Omega$ \\
            Search $\Omega_i$ in $\Lambda$ \\
            %\tcc{match found}
            \If{$\Omega_i.ID == \Lambda_j.CWE\_ID$} 
            {
                Fetch $\Lambda_j.BUS,\Lambda_j.IP,\Lambda_j.TYPE$ \\
                \textbf{\textit{Map\_CWE($\Omega_i$,$\Lambda_j$,$S.IP$,$\omega$)}}
            }
            \ElseIf{$Cosine\_Sim(\Omega_i.DESC,\Lambda_j.DESC) > 0.75$}
            {
                Fetch $\Lambda_j.CWE\_ID,\Lambda_j.BUS,\Lambda_j.IP,$\\
                $\Lambda_j.TYPE$ \\
                Update $\Omega_i.ID = \Lambda_j.CWE\_ID$ \\
                \textbf{\textit{Map\_CWE($\Omega_i$,$\Lambda_j$,$S.IP$,$\omega$)}}
               
            }
            \Else
            {
                Generate Query for LLMs: \\
                `Is $<\Omega_i.ID,\Omega_i.DESC>$ relevant for bus-based Soc with $<SoC\_Config>$'
                
                \If{ $is\_relevant(response)$ }
                {
                    $\Lambda.append(\Omega_i)$ \\
                    $\omega.append(\Omega_i)$
                }
               
            }
       }
       return $\omega$
   %}
\end{algorithm}

\begin{algorithm}[!ht]
\caption{Map CWEs}
\label{algo:map_cwe}

\DontPrintSemicolon
  
% \KwInput{List of CWEs from ChatGPT Responses, denoted by $\Omega$}
% \KwOutput{Filtered List of CWEs, denoted by $\omega$ }
\SetKwFunction{FFind}{Map\_CWE}
%\SetKwFunction{FMain}{Filtering}
 
\SetKwProg{Fn}{Function}{:}{\KwRet}
  \Fn{\FFind{$\Omega_i$,$\Lambda_j$,$IP$,$\omega$}}{
        \If{$\Lambda_j.BUS == $`YES'}
                {
                    \If{$\Omega_i.ID \notin \omega $}
                    {
                        $\omega.append(\Omega_i)$
                    }
                }
                 \If{$\Lambda_j.IP == $`YES'}
                {
                    \If{$\Lambda_j.MISC.IP\_NAME \in IP_k.NAME, \forall k, k \in [1,q]$}
                    {
                         \If{$\Omega_i.ID \notin \omega$}
                          {
                              $\omega.append(\Omega_i)$
                           }
                    }
                    \If{$\Lambda_j.MISC.IP\_TYPE \in IP_k.OP, \forall k, k \in [1,q]$}
                    {
                         \If{$\Omega_i.ID \notin \omega $}
                          {
                              $\omega.append(\Omega_i)$
                           }
                    }
                   
                }
        return $\omega$
  }
\end{algorithm}

    \subsection{Security Assertion Creation \& Verification}     
        After acquiring the list of relevant CWEs, \divas~initiates the creation of security assertions to ascertain  whether the existing SoC implementation is susceptible to these particular vulnerabilities. 
        This step holds significant importance, given that certain vulnerabilities have the potential to result in potential attacks and subsequently breach the CIA properties of a system. Assertion-based verification is widely popular in SoC design flow for detecting any design behavior that deviates from the defined security requirements and thus leads to potential attacks. SVAs are broadly adopted in various phases of the SoC design flow, from specification to verification stages. Security experts or architects are mostly responsible for writing assertions manually and verifying the design behavior. In this work, we have explored the capability of both ChatGPT and BARD to generate SVAs for each of the identified CWEs and the respective IPs of the SoC design. The automated flow involves LLMs for generating initial SVAs and then modifying them using the available design specifications, making it syntactically correct to verify the design under test. All the assertions are generated and appended to the respective IP module, followed by verification through simulation using standard simulators e.g. Synopsys VCS. 

\begin{figure}[h]
\centering
\subfloat[][ChatGPT]{\includegraphics[scale=0.28]{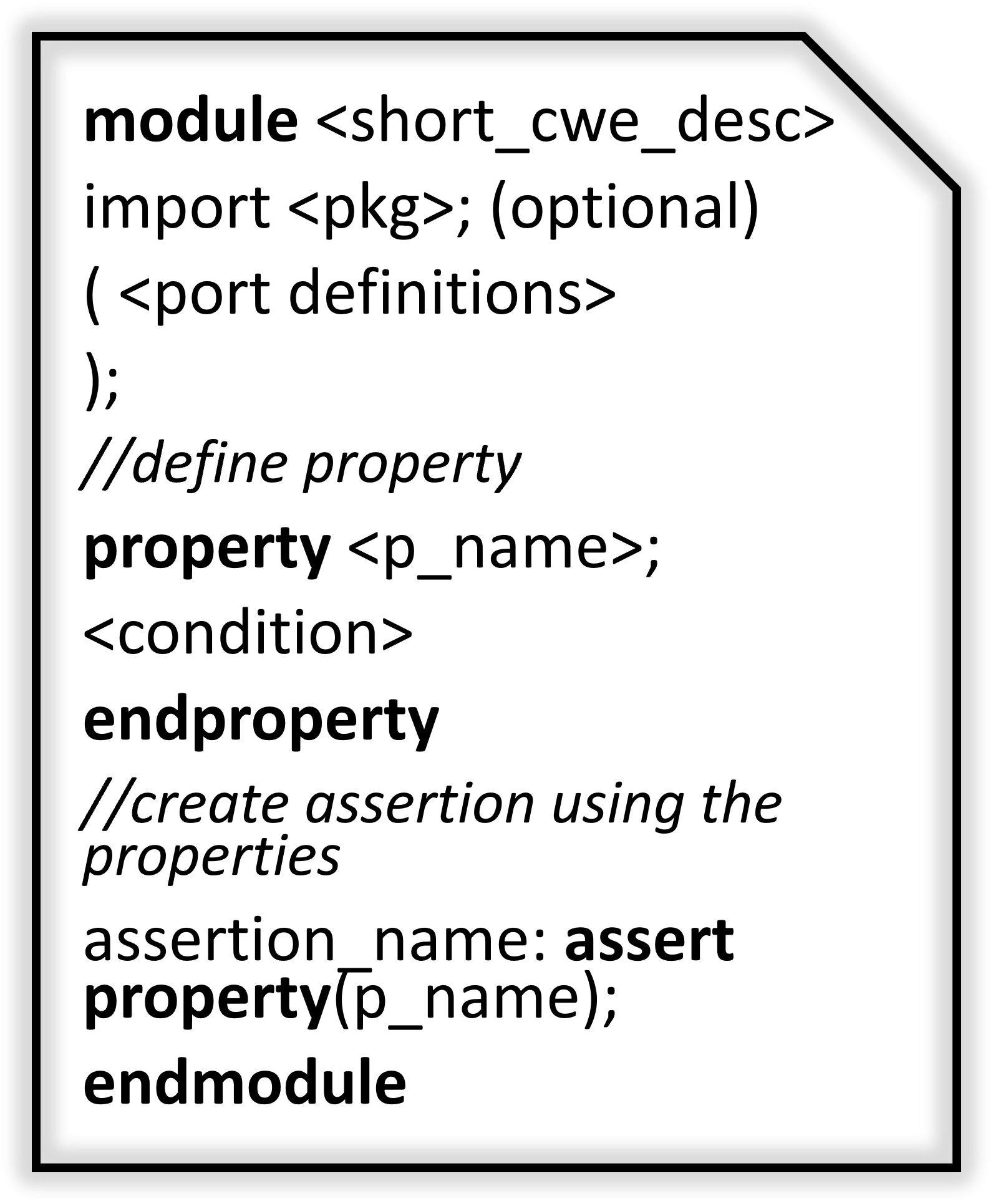}}
\hfill
\subfloat[][BARD]{\includegraphics[scale=0.28]{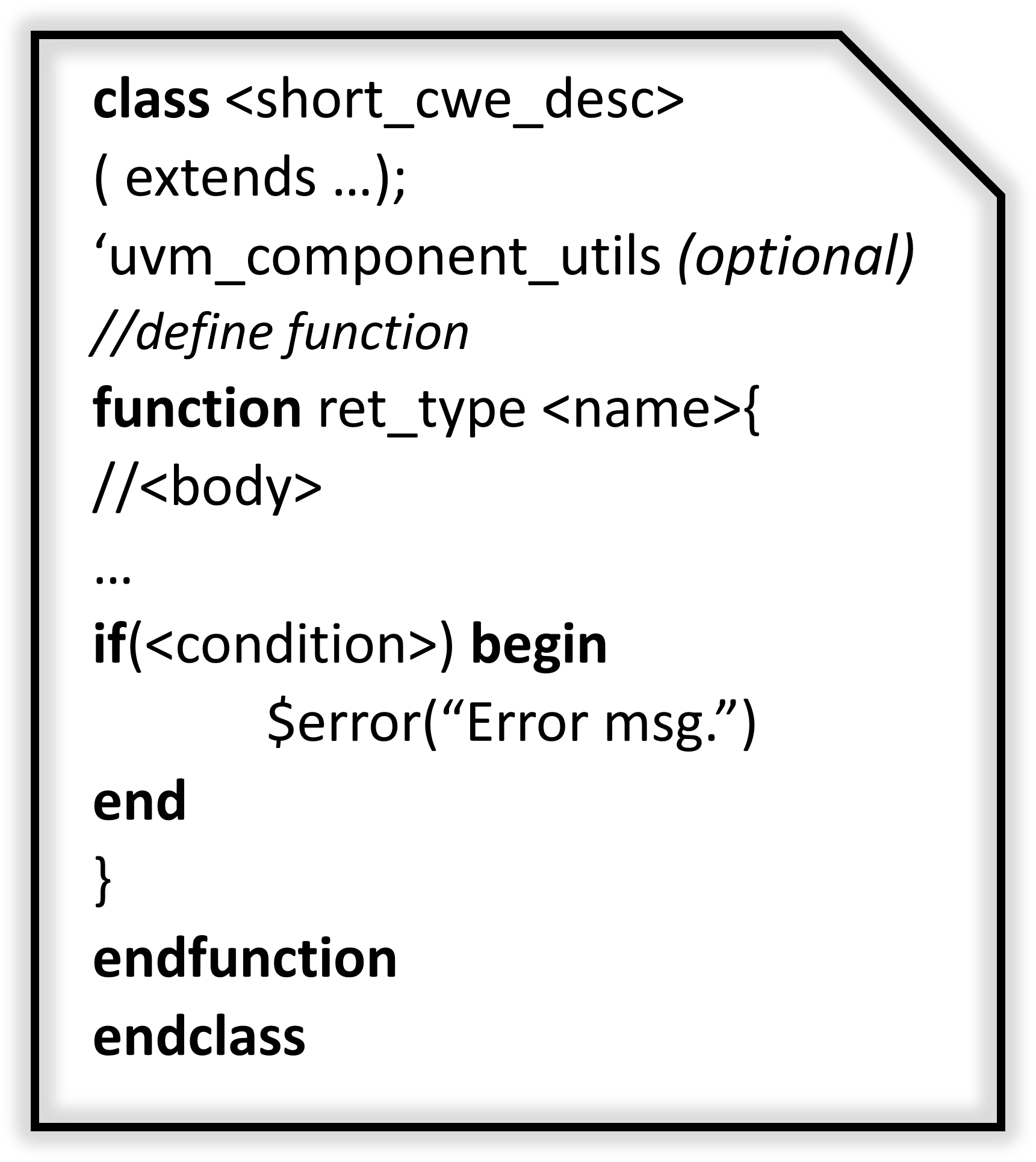}}
 \hfill
\caption{Generalized Assertion Templates.}
\label{fig:assertion_template}
\end{figure}

\begin{table*}[t!]
\centering
\caption{Examples of CWEs and SVAs for Different Types of Security Policies along with the Bug Description}
\label{tab:cwe_to_assertions}
%\resizebox{\columnwidth}{!}{%
\begin{tabular}{|c|c|c|l|}
\hline
\textbf{Type} &
  \textbf{CWE-ID} &
  \multicolumn{1}{c|}{\textbf{Bug Description}} &
  \multicolumn{1}{c|}{\textbf{SystemVerilog Assertion}}
  \\ \hline
Access Control &
  CWE-125 &
  Out-of-bounds Read &
  \begin{tabular}[c]{@{}l@{}}property p\_out\_of\_bounds\_read;\\
$@(posedge(clk\_i))$\$rose(start)$|->$ \\ $(wb\_adr\_i >= 32'h93000004$ \&\&  $wb\_adr\_i$ \\ $<= 32'h93000008);$\\
endproperty \\
a\_out\_of\_bounds\_read: \\
assert property (p\_out\_of\_bounds\_read)\\
else  \$display(``Out-of-bounds Access!'');\end{tabular}
   \\ \hline
\multicolumn{1}{|c|}{Information Flow} &
  CWE-327 &
  \begin{tabular}[c]{@{}c@{}}Use of a Broken or Risky \\ Cryptographic Algorithm\end{tabular} &

  \begin{tabular}[c]{@{}l@{}}property p\_broken\_algo;\\
    $@(posedge(clk\_i))$ \$rose(start) \&\& $key == $\\
    $32'hABCD1234;$\\
        endproperty \\
    a\_broken\_algo: assert property (p\_broken\_algo) \\
    else  \$error(``Key has been left at a default value.''); \end{tabular}
   \\ \hline
   \multicolumn{1}{|c|}{Liveness} &
  CWE-226 &
  \begin{tabular}[c]{@{}c@{}}Sensitive Information in Resource \\ Not Removed Before Reuse\end{tabular} &

  \begin{tabular}[c]{@{}l@{}}property p\_sensitive\_register\_clear;\\
        $@(posedge(clk\_i))$ (release \&\& !rst\_i) \\
        $|->$ (sensitive\_register === 32'b0);\\
    endproperty \\
a\_sensitive\_register\_clear: \\
assert property (p\_sensitive\_register\_clear)\\
        else \$error(``Violation of sensitive register clear rule'');
  \end{tabular}
   \\ \hline
\multicolumn{1}{|c|}{TOCTOU} &
  CWE-362 &
  \begin{tabular}[c]{@{}c@{}}Concurrent Execution using \\ Shared Resource with Improper \\ Synchronization (`Race Condition') \end{tabular} &

  \begin{tabular}[c]{@{}l@{}}property p\_no\_race\_condition; \\
       $@(posedge(clk\_i))$ (!rst\_i) $|->$ \\ !(Slave\_A\_access \&\& Slave\_B\_access); \\
    endproperty\\
a\_no\_race\_condition: \\
assert property (p\_no\_race\_condition)\\
        else \$error(``Violation of no race condition rule'');
  \end{tabular}
   \\ \hline
\end{tabular}%
%}
\end{table*}   

        Fig. \ref{fig:assertion_template} depicts the generalized structure of an assertion created by ChatGPT and BARD, respectively. As evident from the figures, ChatGPT consistently tries to create a module with property with the underlying condition for the assertion, while BARD is inconsistent in generating such a well-formatted template for assertions. BARD attempts to utilize UVM-based packages and functions by extending a class from a superclass which is more specific to the SoC design architecture and has limitations in terms of extensibility and usability. Hence, we opted for the template as depicted in Fig. \ref{fig:assertion_template}(a), which utilizes the property block to define the condition and assert the property for verification purposes. We have tabulated some example CWEs inferring to different types of Security Policies and their respective SVA with proper templates in Table \ref{tab:cwe_to_assertions}. 
        
        The SVAs are verifiable in the pre-silicon testing through testbenches to identify any deviation from the security requirements that might lead to potential attacks. Suppose any assertion is evaluated to be \textit{true} during simulation using open-source or commercial simulation tools, such as Synopsys VCS, ModelSim, etc. In that case, there is a requirement to employ necessary preventive measures through appropriate checks and assignment of signal values as required.

\begin{algorithm}[h!]
\caption{Assertion to Security Policy}
\label{algo:translation}
     
\DontPrintSemicolon
      
\KwInput{Set of Assertions, denoted by $\Upsilon$\\
where, $\Upsilon = \{\upsilon_1,\upsilon_2,...\upsilon_n\}$}
\KwOutput{Set of Security Policies, denoted by $\Phi$ \\
where, $\Phi = \{\phi,\phi_2,...\phi_n\}$}
\For{each $\upsilon \in \Upsilon$}
{  
    \If{$\upsilon.clocking\_block$ != NULL}
    {
        $\phi.mode$ = $<\upsilon.clocking\_block:clock,$\\
        $\upsilon.clocking\_block:reset>$
    }
    \Else
    {
        $\phi.mode = 0$  //default value
    }
    \If{$is\_sequential(\upsilon.property)$}  
    {
         /* Sequential event */\\
         $\psi$ = split$(\upsilon.property,|->)$  \\
         \For{each $\psi_i \in \psi$}
         {
            $\tau$ = fetch\_cycles($\psi_i$:Cycles) \\
            \If{$\tau$ is NULL}
            {
                $\tau = 1$ //default value
            }
            $\psi.predicate$.append($\psi_i$)  \\
            $\psi.predicate$.append(\#\# $\tau$ \#\#) \\
         }  
    }
    \Else 
    {
        /* Boolean Expression */ \\
        $\psi.predicate$ = $\upsilon.property$
    }
    $\psi.action$ = extract\_action(user\_input) 
}
return $\Phi$
\end{algorithm}

\begin{figure*}[b]
\centering
\includegraphics[scale=0.45]{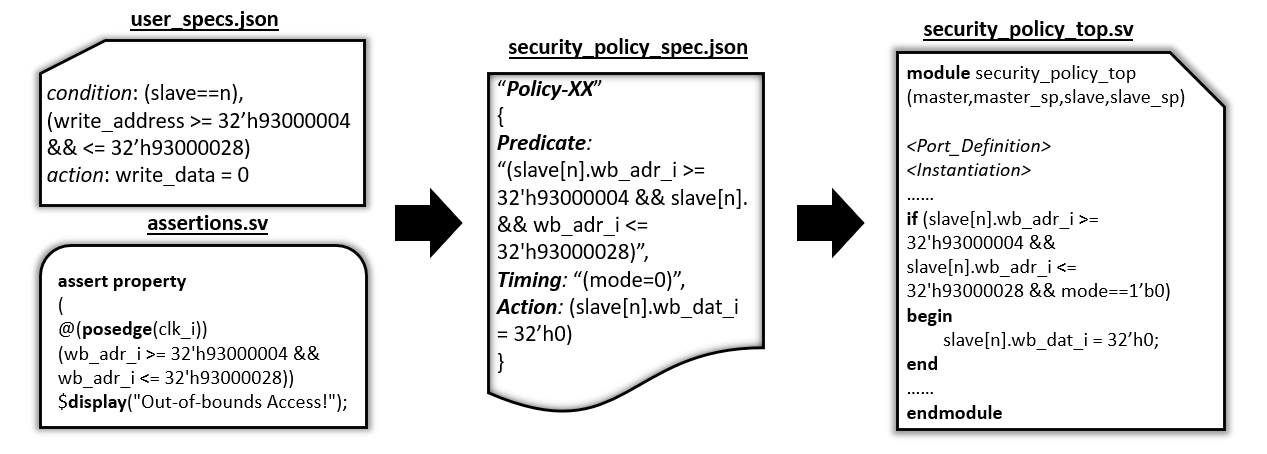}
\caption{Conversion of SystemVerilog Assertion to Security Policy followed by Policy Enforcement.}
\label{assertion_policy_code_fix}
\end{figure*}
        
    \subsection{Translation to Security Policy \& Policy Enforcement}    
    
     Each assertion that is found to be activated during the simulation indicates the presence of respective vulnerabilities in the SoC design. We have employed an automated translation from assertion to corresponding security policy for representing the security requirements formally. Each security policy is expressed in a 3-tuple {\em $<$predicate, timing, action$>$} format. The assertions are formed to check a specific event, represented in the {\em `property...endproperty'} block, which can either be a sequence of events or a static condition-based check. For sequential events, the number of clock cycles between the events is included in the security policy accordingly, while Boolean expression can be used for fixed conditional checks. \divas~is designed to accommodate both scenarios while generating the {\em `predicate'} of a security policy. Additionally, a separate clocking block could be present in the assertion to fulfill specific requirements of {\em `clock'} or {\em `reset'} values. The automated flow includes these values in the {\em `timing'} section of the security policy and any additional clock cycle requirements. It also appends the operating mode represented using an integer value (e.g., for user mode: 0, debug mode: 1, etc.) in the {\em `timing'} tuple if available. The {\em `action'} can be customized based on security requirements for the SoC design. The automated flow is designed to parse the user responses and includes the respective action represented using the assignment of required values for observable signals. Algorithm \ref{algo:translation} includes all the steps for translating each assertion to the respective Security Policy represented in a formal 3-tuple format.

     The automated translation from security policy to Verilog code through DiSPEL framework \cite{dispel_arxiv} has been integrated with this workflow for incorporating policy enforcement in the SoC under consideration. The policies are converted to equivalent Boolean expressions with conditional statements, and the assignment of required values is defined inside the respective conditional blocks. The bus-level policies are incorporated through a centralized security module, whereas the IP-level security policies are appended in each IP module's respective top module wrapper. The updated code with policy enforcement logic can be synthesized using any industry standard tool package to generate a gate-level netlist. We have also employed functional verification using simulation to verify the correctness of each IP module before and after the inclusion of security policies. Fig. \ref{assertion_policy_code_fix} depicts the conversion from SVA to the respective security policy represented in a 3-tuple format followed by the automatic generation of RTL code for policy enforcement using DiSPEL tool \cite{dispel_arxiv}.

\section{Experimental Results}

This section describes the experimental setup and analysis of the obtained results for evaluating the performance of the proposed framework. We have opted for the open-source Common Evaluation Framework (CEP) benchmark suite from MIT-LL \cite{MITCEP} for our experiments. This SoC benchmark consists of a Mor1kx OpenRISC processor as the Master IP and several Slave IPs with different functionalities. The Slave IPs include multiple cryptographic modules, namely AES, DES3, MD5, RSA, and SHA-256, and several digital signal-processing modules such as DFT, IDFT, FIR, IIR, and GPS. Additional modules, like JTAG, GPIO, UART, etc., are used for external communication with the user that could be an attack surface to launch an attack and are considered untrusted IPs.

\begin{figure}[h]
\centering
\includegraphics[scale=0.32]{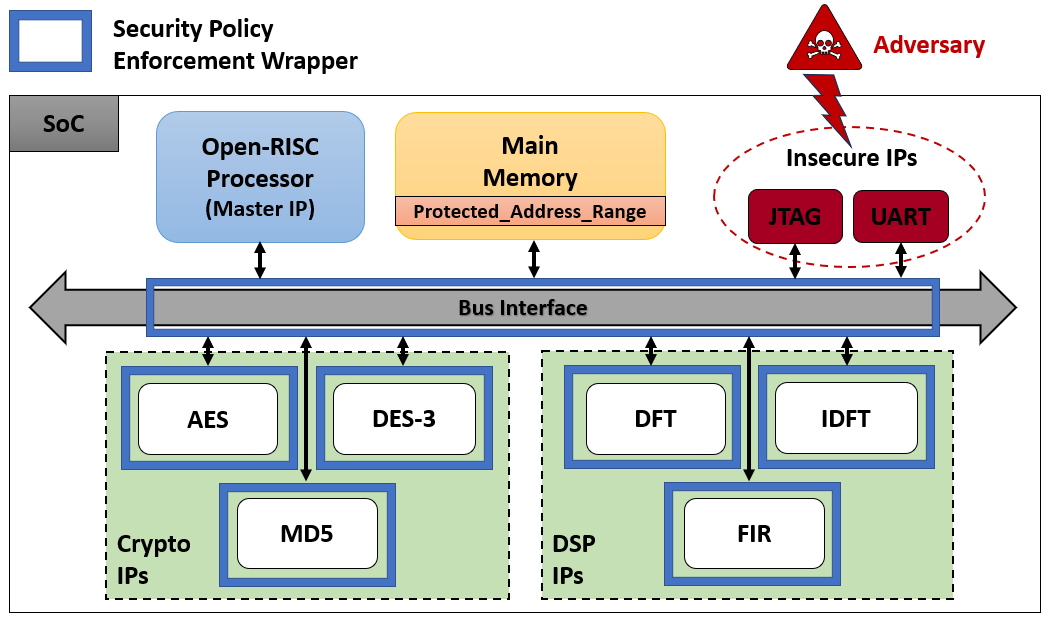}
\caption{The Security Policy Enforcement Wrapper placement in the SoC architecture.}
\label{fig:sample_soc_arch}
\end{figure}

Fig. \ref{fig:sample_soc_arch} illustrates a bus-based SoC architecture consisting of Master IP, Main Memory, Crypto IPs, DSP IPs, and insecure IPs such as JTAG and UART. The centralized security policy enforcement module is placed across the bus interface where all the IPs communicate with other IPs using the bus while the IP-level security enforcement logic is realized through a wrapper above the top level of each IP. To formalize the design specifications of the SoC, we have created a JSON-based template that provides a structured representation of the SoC's design, encompassing the overall configuration and IP-level details. Using this template, we can capture and organize the pertinent information related to the SoC under examination, as depicted in Listing \ref{lst:2}.\\

\lstset{style=custom}
\begin{lstlisting}[caption=SoC Design Specification Example, label=lst:2, language=Java]
{   
    "SoC":
    {
        "NAME":"MIT-CEP",
        "TYPE":"Open-source",
        "USAGE":"Academic Research",
        "BUS":"AXI4",
        "NO_OF_MASTERS":"1",
        "NO_OF_SLAVES":"11"
    },
    "BUS_INTERFACE":
    {
        "INTERFACE_NAME":"Master/Slave",
        "NO_OF_PORTS":"17",
        "SIGNAL_NAMES":"AWVALID,AWADDR,WDATA,ARREADY,RDATA,..."
    },
    "MASTER_1":
    {
    "NAME":"mor1kx",
    "TYPE":"OPen RISC-V",
    "OPERATION":"Processor",
    "ADDRESS_RANGE":"90000000-99000000",
    "BASE_ADDRESS":"90000000",
    "PROTECTED_ADDRESS_RANGE":"9100001F:9100002D"
    },
    "SLAVE_1":
    {
    "NAME":"Advanced Encryption Standard",
    "ABBREVIATION":"AES",
    "TYPE":"Open-source",
    "OPERATION":"Crypto",
    "ADDRESS_RANGE":"93000000-93FFFFFF",
    "BASE_ADDRESS":"93000000",
    "PROTECTED_ADDRESS_RANGE":"93000014:9300003C"
    }
}
\end{lstlisting}

\begin{table*}[t]
\centering
\caption{Generating CWEs under different assumptions}
\label{tab:no_assumption_vs_with_assumptions}
\begin{tabular}{lccc}
\hline
\multicolumn{1}{c}{\multirow{2}*{\textbf{LLM Response under no assumptions}}}  & \multicolumn{3}{c}{\textbf{Assumptions}} \\
\cline{2-4}
 & \textbf{\#1}    & \textbf{\#2} & \textbf{\#3} \\
\hline
CWE-20: Improper   Input Validation                                                           & CWE-20         & CWE-20         & CWE-20         \\
CWE-79: Improper   Neutralization of Input During Web Page Generation & CWE-79         & CWE-79         & CWE-79         \\
CWE-120: Buffer Copy   without Checking Size of Input  & CWE-120        & CWE-120        & CWE-120        \\
CWE-125:   Out-of-bounds Read                                                                 & CWE-134        & CWE-125        & CWE-125        \\
CWE-134: Use of   Externally-Controlled Format String                                         & CWE-284        & CWE-134        & CWE-134        \\
CWE-200: Exposure of   Sensitive Information to an Unauthorized Actor                         & CWE-287        & CWE-200        & CWE-284        \\
CWE-284: Improper   Access Control                                                            & CWE-306        & CWE-319        & CWE-287        \\
CWE-287: Improper   Authentication                                                            & CWE-319        & CWE-330        & CWE-306        \\
CWE-306: Missing   Authentication for Critical Function                                       & CWE-330        & CWE-400        & CWE-400        \\
CWE-319: Cleartext   Transmission of Sensitive Information                                    & CWE-400        & CWE-416        & CWE-416        \\
CWE-330: Use of   Insufficiently Random Values                                                & CWE-416        & CWE-787        & CWE-732        \\
CWE-400: Uncontrolled   Resource Consumption                                                  & CWE-732        & CWE-798        & CWE-787        \\
CWE-416: Use After   Free                                                                     & CWE-798        &                & CWE-798        \\
CWE-732: Incorrect   Permission Assignment for Critical Resource                              & CWE-862        &                & CWE-862        \\
CWE-787:   Out-of-bounds Write                                                                &                &                &                \\
CWE-798: Use of   Hard-coded Credentials                                                      &                &                &                \\
CWE-862: Missing   Authorization                                                              &                &                &       \\  
\hline     
\end{tabular}
\end{table*}

\subsection{Generating CWEs using LLMs}

Both ChatGPT and BARD are found to be very powerful and versatile LLMs due to their advanced architecture, large-scale training data, fine-tuning capabilities, and ongoing research. We have generated context-specific questions from SoC design specifications and explored the capabilities of ChatGPT and BARD models to generate relevant CWEs for a given SoC design. The following block depicts an example scenario of generating CWEs for a given SoC configuration using ChatGPT (GPT-4 model).

\begin{tcolorbox}[breakable,pad at break*=1mm,colback=gray!5!white,colframe=black!50!black,
colbacktitle=gray!55!black,title= Q1: Find the below SoC Configuration: \\
$<SoC\_Config>$\\
Can you generate all possible relevant CWEs for this SoC configuration?]
\begin{itemize}
   \item CWE-20: Improper Input Validation
    \item CWE-79: Improper Neutralization of Input During Web Page Generation
    \item CWE-89: Improper Neutralization of Special Elements used in an SQL Command
    \item CWE-120: Buffer Copy without Checking Size of Input
    \item CWE-134: Use of Externally-Controlled Format String
    \item CWE-190: Integer Overflow or Wraparound
    \item CWE-287: Improper Authentication
    \item CWE-306: Missing Authentication for Critical Function
    \item CWE-311: Missing Encryption of Sensitive Data
    \item CWE-319: Cleartext Transmission of Sensitive Information
    \item CWE-330: Use of Insufficiently Random Values
    \item CWE-399: Resource Management Errors
    \item CWE-400: Uncontrolled Resource Consumption
    \item CWE-416: Use After Free
    \item CWE-426: Untrusted Search Path
    \item CWE-601: URL Redirection to Untrusted Site
    \item CWE-732: Incorrect Permission Assignment for Critical Resource
    \item CWE-759: Use of a One-Way Hash without a Salt
    \item CWE-798: Use of Hard-coded Credentials
    \item CWE-862: Missing Authorization
    \item CWE-918: Server-Side Request Forgery
\end{itemize}
\end{tcolorbox}

We have also explored how the LLM response during CWE generation can vary under different assumptions for identical SoC configurations. We observed that the CWEs listed by a given LLM get refined based on the assumption. However, this implied that the list of CWEs generated under any assumption might not be comprehensive due to the LLM's limited knowledge base and lack of contextual understanding. Hence, we need a filtration step after query generation to identify the relevant CWEs from the list. Table \ref{tab:no_assumption_vs_with_assumptions} shows the variations in the list of CWEs generated by GPT-4 under different assumption settings:
\begin{itemize}
    \item \textit{No Assumptions}.
    \item \textit{Assumption \#1:} Bus Transactions are secure.
    \item \textit{Assumption \#2:} Proper Authentication \& Authorizations are in place.
    \item \textit{Assumption \#3:} Side Channel Attacks are infeasible.\\
\end{itemize}

\begin{table*}[t]
\centering
\caption{Unfiltered vs. Filtered List of CWEs}
\label{tab:filtering}
\begin{tabular}{|>{\centering\arraybackslash}p{7.5cm}|>{\centering\arraybackslash}p{1.25cm}|>{\centering\arraybackslash}p{8cm}|}
%{|p{0.35\linewidth}|p{0.08\linewidth}|p{0.5\linewidth}|}
\hline
\textbf{Unfiltered   List of CWEs} & \textbf{Relevant?} & \textbf{Filtered List of Relevant CWEs \& Classification}                                                                                                                                       \\ \hline
CWE-119: Improper Restriction of Operations within the Bounds of a Memory Buffer
& \multirow{2}{*}{\textbf{No}}
& \multirow{11}*{
\begin{tabular}{>{\centering\arraybackslash}p{7cm}}\\
%{c}
\{`CWE-200', `Information Exposure', `Bus, IP', `Sync', `Access Control'\}\\
\{`CWE-284', `Improper Access Control', `Bus, IP', `Async', `Access Control'\}\\
\{`CWE-310', `Cryptographic Issues (e.g., weak algorithms or insecure implementation)', `IP', `Async', `Information Flow'\}\\
\{`CWE-325', `Missing Required Cryptographic Step, `IP', `Async', `Information Flow'\}\\
\{`CWE-261', `Weak Cryptography for Passwords', `IP', `Async', `Information Flow'\}\\
\{`CWE-362', `Concurrent Execution using Shared Resource with Improper Synchronization (`Race Condition')', `Bus', `Sync', `Liveness'\}
\end{tabular}} \\ 

\cline{1-2}
CWE-20: Improper Input Validation
& \multirow{1}{*}{\textbf{No}}                  
& \\ 
\cline{1-2}
CWE-200: Information Exposure
& \cellcolor{green!25}{\textbf{Yes}}                
& \\ 
\cline{1-2}
CWE-352: Cross-Site Request Forgery (CSRF)
& \multirow{1}{*}{\textbf{No}}                 
& \\ 
\cline{1-2}
CWE-284: Improper Access Control
& \cellcolor{green!25}\multirow{1}{*}{\textbf{Yes}}                  
& \\ 
\cline{1-2}
CWE-94: Improper Control of Generation of Code ('Code Injection')
& \multirow{2}{*}{\textbf{No}}              
& \\ 
\cline{1-2}
CWE-377: Insecure Temporary File 
& \multirow{1}{*}{\textbf{No}}                
& \\ 
\cline{1-2}
CWE-310: Cryptographic Issues (e.g., weak algorithms or insecure implementation)
& \cellcolor{green!25}\multirow{2}{*}{\textbf{Yes}}
& \\ 
\cline{1-2}
CWE-325: Missing Required Cryptographic Step
& \cellcolor{green!25}\multirow{1}{*}{\textbf{Yes}}
& \\ 
\cline{1-2}
CWE-261: Weak Cryptography for Passwords
& \cellcolor{green!25}\multirow{1}{*}{\textbf{Yes}}
& \\ 
\cline{1-2}
CWE-362: Concurrent Execution using Shared Resource with Improper Synchronization (`Race Condition') 
& \cellcolor{green!25}\multirow{3}{*}{\textbf{Yes}}  
& \\ 
\hline
\end{tabular}
\end{table*}

\noindent \textbf{Limitations of LLMs while identifying CWEs}

During our experiments for generating relevant CWEs for a given SoC, we observed that the LLMs have several limitations. Some of these limitations are listed as follows:
    \begin{enumerate}[i.]
        \item Inconsistent Ranking of CWEs. 
        \item Overlapping results for different IPs with similar functionality.
        \item Incorrect mapping between CWE-ID and the Bug description.
        \item Generating out-of-context CWEs on multiple occasions.
        \item Ambiguity between description and explanation for the same CWE-ID.
        \item Mapping multiple issues to the same CWE-ID.
        \item Incomplete list of CWEs for any particular configuration under different assumptions.
    \end{enumerate}

\subsection{Filtering CWEs}
\begin{figure}[!ht]
\centering
\includegraphics[scale=0.5]{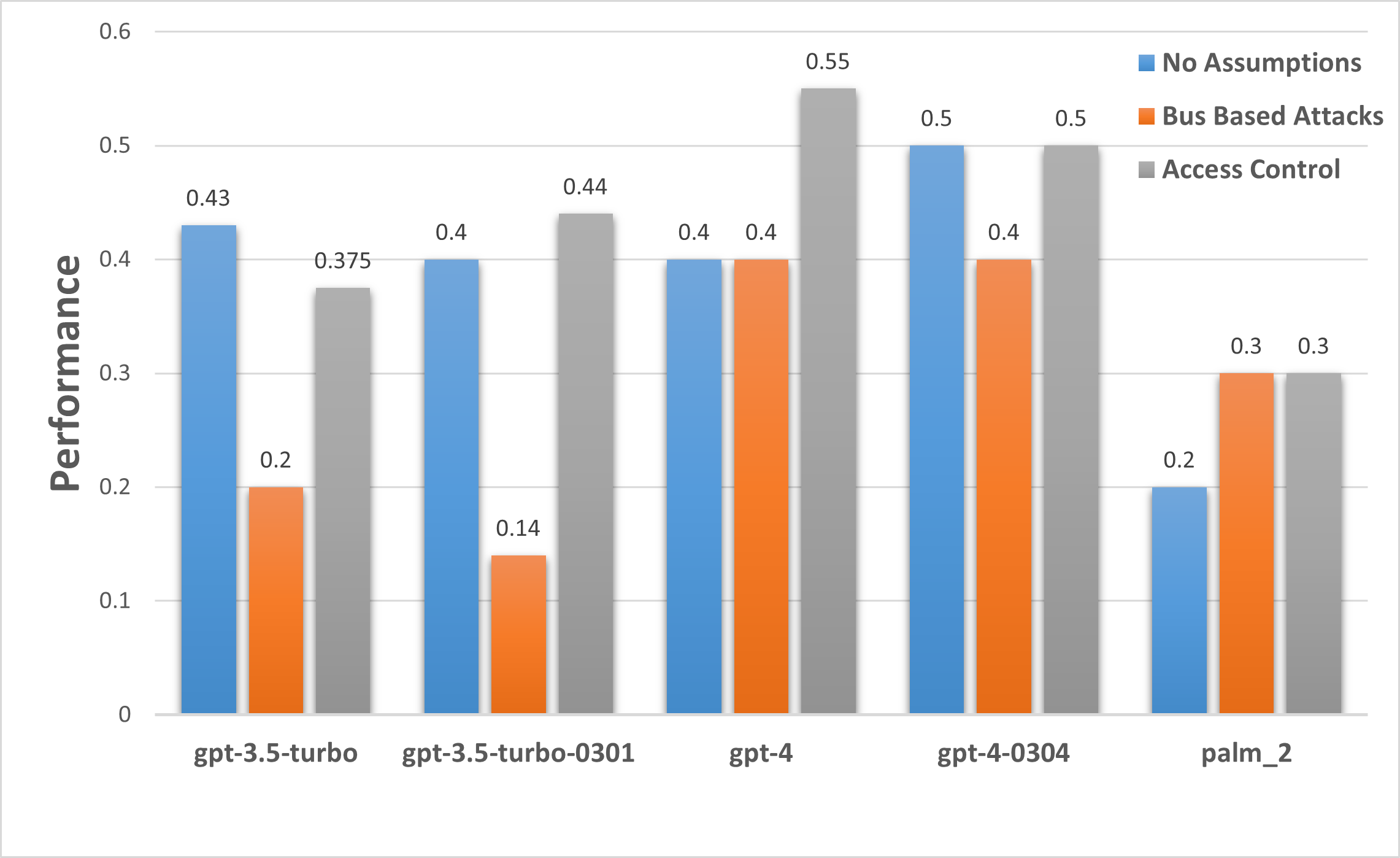}
\caption{Performance analysis of LLMs in generating relevant CWEs for a given SoC configuration.}
\label{fig:bar_graph_performance_2}
\end{figure}

A filtering mechanism is required to identify only relevant CWEs and discard non-relevant or out-of-context CWEs for a given SoC configuration. The filtering on LLM response helps reduce any bias or ambiguity, maintain consistency, correctly map CWE-ID to respective descriptions, and improve accuracy for better performance. Table \ref{tab:filtering} lists the CWEs obtained from LLM responses and the filtered list of relevant CWEs for the MIT-CEP SoC. We quantify the performance of an LLM for a given SoC configuration by determining the number of relevant CWEs out of all the CWEs listed in the generated response. The performance metric is calculated for each LLM tested using the \divas~ framework using Equation \ref {eq:1}, and the resulting performance analysis under different assumptions is shown in Fig. \ref{fig:bar_graph_performance_2}.

\begin{equation}
\label{eq:1}
    Performance =  \frac{{\#CWE_{relevant}}}{{\#CWE_{total}}}
\end{equation}

\begin{table*}[!ht]
\centering
\caption{Example Semantic Textual Similarity scores for CWE descriptions}
\label{tab:cosine_score}
\resizebox{\textwidth}{!}{%
\begin{tabular}{|p{0.32\linewidth}|p{0.5\linewidth}|p{0.15\linewidth}|}
\hline
\textbf{Description   Text from LLM Response}          & \multicolumn{1}{c|}{\textbf{Description   Text from Extensive DB (with Score \textgreater 0.4)}} & \textbf{Cosine\_Similarity} \\ \hline
\multirow{4}{*}{Weak   Cryptography for Passwords}     & Cryptographic   Issues                                                                           & Score: 0.5466                      \\ \cline{2-3} 
                                                       & \textcolor{blue}{Inadequate Encryption Strength (Weak key length)}                                                 & Score: \textcolor{blue}{0.5593}                      \\ \cline{2-3} 
                                                       & Use of a Broken or Risky Cryptographic Algorithm                                                 & Score: 0.4929                      \\ \cline{2-3} 
                                                       & Use of Hard-coded Credentials                                                                    & Score: 0.4371                      \\ \hline
\multirow{3}{*}{Missing   Required Cryptographic Step} & Missing   Authentication for Critical Function                                                   & Score: 0.4487                      \\ \cline{2-3} 
                                                       & \textcolor{blue}{Cryptographic Issues}                                                                             & Score: \textcolor{blue}{0.5700}                      \\ \cline{2-3} 
                                                       & Use of a Broken or Risky Cryptographic Algorithm                                                 & Score: 0.4147                      \\ \hline
\end{tabular}%
}
\end{table*}

\subsection{Assertion Generation \& Correction}
We leveraged the capability of LLMs to generate SVAs for a given CWE-ID. The following block illustrates how to generate the SVA for CWE-284 (Improper Access Control). The SVA generated by ChatGPT checks if a specific master (Master A) is trying to access a secure memory region. This example assumes the master can be accessed using the master ID (`$mid$') and the target address using `$addr$'.\\

\begin{tcolorbox}[breakable,pad at break*=1mm,colback=gray!5!white,colframe=black!50!black,
colbacktitle=gray!55!black,title= Q2: Can you generate SystemVerilog Assertion for CWE-284 for a bus based SoC which uses AXI4 bus protocol?]

\lstset{style=custom}
\begin{lstlisting}[caption={}, label={}, language=Verilog]

module improper_access_control_assertion (
    input wire clk,
    input wire rst,
    input wire [31:0] mid,
    input wire [31:0] addr
);

    // Define the clocking and reset behavior
    default clocking cb @(posedge clk);
        default input #1step rst;
    endclocking

    // Let's assume that Master A has an ID of 0, and the secure memory region starts at address 0x80000000
    localparam MASTER_A_ID = 32'b0;
    localparam SECURE_MEMORY_START_ADDR = 32'h80000000;

    // Property: Master A should not access the secure memory region
    property p_improper_access_control;
        @(cb) (!rst) |-> ((mid === MASTER_A_ID) -> (addr < SECURE_MEMORY_START_ADDR));
    endproperty

    // Assertion: Check that the property holds true throughout the simulation
    a_improper_access_control: assert property (p_improper_access_control)
        else $error("Violation of improper access control rule");

endmodule
\end{lstlisting}

\end{tcolorbox}

\noindent \textbf{Limitations of LLMs while generating SVAs}

We have thoroughly tested many scenarios to evaluate the performance of both ChatGPT and BARD in generating SystemVerilog assertions from given CWEs for any IP in any bus-based SoC. Though it works exceptionally well in generating context-related Verilog and SystemVerilog code for the respective security assertions, it falls short in syntactic correctness and occasionally deviates from the actual requirements. The generated description text for assertions seems similarly structured, which may be out-of-context for the given query text. We have also found that the generated texts for sequentially related queries for the same IP are just a paraphrased version of the previous one or sometimes entirely irrelevant in that context. 

We are listing down some of the common shortcomings based on our observations during the experiment as follows:
\begin{enumerate}[i.]
    \item Generating wrong words which sound similar but are not valid keywords for                     SystemVerilog. \\
            %Generating similar words but out of context\\
            \textbf{Example:} `$bit$', `$beats$' in place of `$bits$'
    \item Combining Verilog keywords during assertion generation. \\
            \textbf{Example:} {\em `disable\_iff' or `disable if' or `disable\_if'} (Wrong)\\
            %`$disable\_iff$' or `$disable if$' or `$disable\_if$' (Wrong)\\
                \textbf{Correct:} {\em disable iff}
                %`$disable iff$'
    \item Using contradicting conditions that may disable the assertion during simulation. \\
          \textbf{Example:}
\lstset{style=custom}
\begin{lstlisting}[caption={}, label={}, language=Verilog]
disable iff(wb_sel_i || wb_we_i || !wb_cyc_i)
\end{lstlisting}
            %Ex: disable iff$(wb\_sel\_i || wb\_we\_i || !wb\_cyc\_i)$\\
            The value of {\em wb\_sel\_i} is set as {\em 1'b1} when the corresponding IP is selected, and hence it creates a contradiction during simulation, and the assertion is skipped. 
            %\textcolor{red}{(need to write the logic explanation)}
    \item Missing `$@$' while generating an assertion that causes a syntax error.\\
    \textbf{Example:}
\lstset{style=custom}
\begin{lstlisting}[caption={}, label={}, language=Verilog]
@(posedge(clk_i)) <condition..>
@(negedge (rst_i)) <condition..>
\end{lstlisting}
            %Ex: $@(posedge(clk\_i))$ or $@(negedge (rst\_i))$
            %\textcolor{red}{(write example)}
    \item Invalid ordering of sequential events or missing events while generating the assertion that skip valid checking during simulation. \\
           \textbf{Example:}
\lstset{style=custom}
\begin{lstlisting}[caption={}, label={}, language=Verilog]
@(posedge(clk_i))
$rose(ready && !rst) |-> valid; 
\end{lstlisting}
            The above assertion contains an invalid sequence where {\em `ready'} is checked before the `$valid$' signal is set. The assertion also skips the checking on the {\em `start'} signal. 
                % $@(posedge(clk\_i))$\\
                % \$rose(ready \&\& !rst) $ |-> $ valid; 
            %\textcolor{red}{(write example)}
    \item Contrasts while using the system functions like `{\em \$error', `\$info', `\$display'} in           assertions.\\
          % Ex: The following assertion uses `\$info' instead of `\$error' function, \\ 
          % $@(posedge(clk\_i)) \$rose(start) |-> $ \\
          % \$info(``$<$Error\_Message$>$'')
          %   %\textcolor{red}{(write example)}
          \textbf{Example:}
\lstset{style=custom}
\begin{lstlisting}[caption={}, label={}, language=Verilog]
@(posedge(clk_i)) $rose(start) |-> 
$info("< Error_Message >")
\end{lstlisting}    
        The assertion mentioned above uses {\em `\$info'} instead of {\em `\$error'} function. %\\ $@(posedge(clk\_i)) \$rose(start) |-> $ 
    \item Activating assertion with the wrong error message due to incorrect conditional statements.\\
          \textbf{Example:}
\lstset{style=custom}
\begin{lstlisting}[caption={}, label={}, language=Verilog]
//assertion for CWE-XXX
assertion_name: assert property (!p_name)
else $error("CWE-XXX: <Error Message>");
\end{lstlisting} 
        The assertion in the above example generates an invalid error message due to the wrong construction of the conditional block.
        % Ex: The following assertion generates an invalid error message due to the wrong construction of conditional block \\
        % //assertion for CWE-XXX \\
        % assertion\_name: assert property (p\_name)\\
        % else \$error(``CWE-XXX: $<$Error Message$>$'');
        %\textcolor{red}{(write example)}
    
\end{enumerate}

\subsection{Overhead Analysis after Security Policy Enforcement}

The IPs were synthesized using the LEDA 250nm standard cell library from Synopsys in Synopsys Design Compiler to obtain representative overhead values. Table \ref{tab:overhead_table} represents the additional overhead incurred in terms of area, power, and delay after incorporating security policies in the centralized module and particular IPs. The results indicate that the overheads are mostly minimal under default synthesis settings without any constraints. In some scenarios, the overall overheads are reduced after re-synthesizing with security policy module implementation due to internal (heuristic-based) optimizations and hence reported as negligible. Hence, we can conclude that our proposed methodology of enforcing security requirements through policies for a generic bus-based SoC design incurs minimal overheads and is practically viable to implement.

\begin{table}[h]
\centering
\begin{threeparttable}
\caption{Overhead Analysis of Different IPs after Security Policy Enforcement}
\label{tab:overhead_table}
\begin{tabular}{ccccc}
\hline
\multirow{2}*{\textbf{IP}} & \multirow{2}*{\textbf{\#Policies}} & \multicolumn{3}{c}{\textbf{Synthesis Overheads (\%)}} \\
\cline{3-5}
 & & \textbf{Area}    & \textbf{Delay} & \textbf{Power} \\
\hline
AES & 5 & $0.19\uparrow$ & $-3.55\downarrow$ & $ 41.57\uparrow $\\ 
DES3 & 4 & $8.03\uparrow$ & $-2.05\downarrow$ & $10.69\uparrow $\\ 
SHA256 & 4 & $2.73\uparrow$ & $4.39\uparrow$ & $-2.03\downarrow $\\ 
MD5 & 3 & $3.33\uparrow$ & $8.47\uparrow$ & $-6.63\downarrow $\\ 
Main Memory & 2 & $-0.021\downarrow$ & $-0.192\downarrow$ & $-0.079\downarrow $\\ 
\hline
\end{tabular}
\end{threeparttable}
\end{table}

\subsection{Discussion}

Identifying the relevant CWEs for a specific SoC configuration posed a significant challenge due to the lack of accuracy and contextual listing found in both ChatGPT and BARD models. The earlier versions of ChatGPT (v3.x) performed inadequately by generating mostly incorrect responses, lacking in information, and sometimes unrelated to the given SoC context. However, the updated ChatGPT models (v4.x) showed considerable improvement in generating and listing CWEs, although accuracy remained a primary concern. Google BARD can search the web in real-time to find the most recent answers, yet it performs poorly while generating CWEs and assertions for a given configuration. To address this, we implemented a filtering technique to exclude non-relevant CWEs and generate a list of potential vulnerabilities that apply to the given design. Analyzing all the CWEs related to hardware designs listed by the community on the MITRE website enabled us to prepare an extensive database of CWEs with respective classifications, which were then utilized in the filtering process. The filtering step can be evaded if the performance of LLMs in generating relevant CWEs is improved. Current LLMs are trained on diverse texts and topics with fixed domain-specific knowledge, limiting the contextual boundaries of accurate data and detailed information related to SoC security. Incorporating domain-specific LLMs by fine-tuning general-purpose LLMs like ChatGPT or BARD with extensive and accurate domain-specific data would yield better performance. 

\section{Conclusion}
The proposed framework demonstrates that identifying related CWEs for a particular bus-based SoC configuration using LLMs is attainable with some limitations. The inclusion of the filtering process refines the LLM responses to recognize only the relevant CWEs using the Extensive DB of CWEs prepared after meticulously studying and conducting several experiments. The adaption of SVA-based verification using simulation helps to detect the presence of vulnerabilities in the present implementation. The automated translation from assertions to respective security policies paves the way to generate policy enforcement logic either through a centralized module or at the respective IP modules to implement the security requirements. The experimental results demonstrate the security requirements are fulfilled with minimal overhead incurred. The end-to-end automated flow, from identifying potential vulnerabilities to mitigating them by enforcing respective security policies, can be easily integrated and adapted for any standard bus-based SoC configuration, which helps reduce the manual efforts by security experts and provides more flexibility and controllability to the end user. While the current study focuses on bus-based SoCs, the methodology can be extended to other interconnect fabrics, including Network-on-Chip (NoC) based SoCs.

%\section*{Acknowledgments}

%\newpage

%\section*{Biography Section}

\bibliographystyle{IEEEtran}
\bibliography{IEEEabrv, references}

\vfill

\end{document}